\documentstyle[twoside,fleqn,espcrc2,psfig]{article}

\newcommand{\AmS}{{\protect\the\textfont2
  A\kern-.1667em\lower.5ex\hbox{M}\kern-.125emS}}
\newcommand{\be}{\begin{equation}}
\newcommand{\ee}{\end{equation}}
\newcommand{\ben}{\begin{eqnarray}}
\newcommand{\een}{\end{eqnarray}}

\def\simgt{\rlap{\lower 3.5 pt\hbox{$\mathchar \sim$}}\raise 1pt \hbox {$>$}}
\def\simlt{\rlap{\lower 3.5 pt\hbox{$\mathchar \sim$}}\raise 1pt \hbox {$<$}}

\title{The Kaon $B$-parameter with the Wilson Quark 
       Action using Chiral Ward Identities
       \thanks{presented by Y.~Kuramashi}}

\author{JLQCD Collaboration \\[1mm] 
        S.~Aoki\address{Institute of Physics, University of
        Tsukuba, Tsukuba, Ibaraki 305, Japan},
        M.~Fukugita\address{Institute for Cosmic Ray Research, 
        University of Tokyo, Tanashi, Tokyo 188, Japan},
        S.~Hashimoto\address{Computing Research Center, 
        High Energy Accelerator Research Organization(KEK), \\
        Tsukuba, Ibaraki 305, Japan},
        N.~Ishizuka$^{\rm a}$,
        Y.~Iwasaki$^{\rm a,}$\address{Center for Computational 
        Physics, University of Tsukuba, Tsukuba, Ibaraki 305, Japan},
        K.~Kanaya$^{\rm a,d}$, 
        Y.~Kuramashi\address{Institute of Particle and Nuclear Studies, 
        High Energy Accelerator Research Organization(KEK), 
        Tsukuba, Ibaraki 305, Japan },
        M.~Okawa$^{\rm e}$, A.~Ukawa$^{\rm a}$,
        T.~Yoshi\'{e}$^{\rm a,d}$}

\begin{document}

\begin{abstract}

We present a detailed description of the method and results of our  
calculation of 
the kaon $B$ parameter using the
Wilson quark action in quenched QCD at $\beta=5.9-6.5$.
The mixing problem
of the $\Delta s=2$ four-quark operators is solved non-perturbatively
with full use of chiral Ward identities.
We find $B_K({\rm NDR}, 2{\rm GeV})=0.562(64)$ in the continuum limit,
which agrees with the value obtained with the Kogut-Susskind 
quark action.  

\end{abstract}

\maketitle

\section{Introduction}

The Wilson quark action explicitly breaks chiral symmetry
at finite lattice spacing, which causes problems in a 
number of subjects treated by numerical simulations of lattice QCD.
For the calculation of the kaon $B$ parameter $B_K$, the problem 
appears as a non-trivial mixing of the weak $\Delta s=2$ four-quark operator 
of purely left handed chirality with those of mixed left-right 
chirality.  
It has been well known that perturbation theory does
not work effectively for solving this mixing problem\cite{latt88}, and 
most calculations of $B_K$
have tried to resolve the mixing non-perturbatively 
with the aid of chiral perturbation theory\cite{bk_w}.
This method,
however, has not been successful, since it contains large
systematic uncertainties from higher order effects of chiral perturbation 
theory which survive even in the continuum limit.

An essential step for a precise determination of $B_K$ 
is to control the operator mixing non-perturbatively
without resort to any effective theories.
The failure of the perturbative approach suggests that
higher order corrections in terms of the coupling
constant might be large in the mixing coefficients.  Presence of 
large $O(a)$ corrections in the coefficients is also a possibility.
In order to deal with this problem, the Rome group has proposed the
method of non-perturbative renormalization(NPR)\cite{npr}, which shows
an improvement of the chiral behavior for the $\Delta s=2$
operator\cite{romeBK}.
Recently we have proposed an alternative non-perturbative method 
to solve the operator mixing problem which is based on the use of 
chiral Ward identities\cite{bk_w_jlqcd}.
This method fully incorporates the chiral properties 
of the Wilson action.  Our numerical results show that 
the renormalized $\Delta s=2$ operator constructed with this method 
has good chiral behavior with
the mixing coefficients having small momentum scale dependence. 

The chief findings of our calculation have already been presented 
in Ref.~\cite{bk_w_jlqcd}.
In this report we give a detailed description of 
the implementation of our method and the results of our analyses.

\section{Formulation of the method}

Let us consider flavor $SU(3)$ chiral variation defined by
\ben
\delta^a \psi(x)&=&i\frac{\lambda^a}{2}\gamma_5\psi(x), \\
\delta^a {\bar \psi}(x)&=&{\bar \psi}(x)i\frac{\lambda^a}{2}\gamma_5, 
\een
where $\lambda^a$ $(a=1,\cdots,8)$ are the flavor matrices
normalized as Tr$(\lambda^a\lambda^b)=2\delta^{ab}$.
We consider a set of weak operators in the continuum  
$\{\hat O_i\}$ which closes under flavor 
chiral rotations 
$\delta^a \hat O_i=ic^a_{ij}\hat O_j$.  These operators are
given by linear combinations of a set of lattice local operators
$\{O_\alpha\}$ as $\hat O_i=\sum_\alpha Z_{i\alpha}O_\alpha$.

We choose the mixing coefficients $Z_{i\alpha}$ such that
the Green functions of $\{\hat O_i\}$ with quarks in
the external states satisfy the chiral
Ward identity to $O(a)$.  This identity can be derived in a
standard manner\cite{wi} and takes the form given by 
\[
-2\rho Z_A^{\rm ext}\langle\sum_xP^a(x)\hat O_i(0)
\prod_k\tilde\psi(p_k)\rangle   
\]
\vspace{-5mm}
\be
+c^a_{ij}\langle\hat O_j(0)\prod_k\tilde\psi(p_k)\rangle 
\label{eq:wi}
\ee
\vspace{-5mm}
\[
-i\sum_l\langle\hat O_i(0)\prod_{k\ne
l}\tilde\psi(p_k)\delta^a\tilde\psi(p_l)\rangle+O(a)=0, 
\]
where $p_k$ is the momentum of the external quark,
$Z_A^{\rm ext}$
and $\rho=(m-\delta m)/Z_A^{\rm ext}$ are constants to be determined from the 
Ward identities for the axial vector currents\cite{rhoza},
and $P^a$ is the pseudoscalar density of flavor $a$ defined by 
$P^a(x)={\bar {\psi(x)}}\frac{1}{2}\lambda^a\gamma_5{\psi(x)}$.
We note that the first term in (3) comes from the chiral variation 
of the Wilson quark action and the third represents the chiral 
rotation of the external fields.

The four-quark operator relevant for $B_K$ is given by 
$\hat O_{VV+AA}=(\bar s\gamma_\mu d)(\bar s\gamma_\mu d)
+(\bar s\gamma_\mu\gamma_5 d)(\bar s\gamma_\mu\gamma_5 d)$  
where $()$ means color trace.
To fix the mixing coefficients for the lattice four-quark 
operators, we may choose a particular $SU(3)$ flavor chiral
rotation to be applied for $\hat O_{VV+AA}$. 
Avoiding flavor rotations that yield operators
which have Penguin contractions and hence mix with lower dimension 
operators,
we employ the $\lambda^3={\rm diag}(1,-1,0)$ chiral rotation,
under which
$\hat O_{VV+AA}$ and  $\hat O_{VA}=(\bar s\gamma_\mu d)
(\bar s\gamma_\mu\gamma_5 d)$ form a minimal closed set of 
the operators.

Since $\hat O_{VV+AA}$ and $\hat O_{VA}$ are dimension six
operators with $\Delta s=2$, we can restrict ourselves to 
dimension six operators for the construction of the lattice
operators corresponding to them.
The set of lattice bare operators with even parity is given by  
\be
\begin{array}{lll}
VV&=&(\bar s\gamma_\mu d)(\bar s\gamma_\mu d) \\ 
AA&=&(\bar s\gamma_\mu\gamma_5 d)(\bar s\gamma_\mu\gamma_5 d) \\  
SS&=&(\bar s d)(\bar s d) \\ 
PP&=&(\bar s\gamma_5 d)(\bar s\gamma_5 d) \\ 
TT&=&(\bar s\sigma_{\mu\nu} d)(\bar s\sigma_{\mu\nu} d)/2 \\ 
\end{array}
\ee
and the set with odd parity is 
\be
\begin{array}{lll}
VA&=&(\bar s\gamma_\mu d)(\bar s\gamma_\mu\gamma_5 d) \\ 
SP&=&(\bar s d)(\bar s\gamma_5 d) \\  
T{\tilde T}&=&(\bar s\sigma_{\mu\nu} d)(\bar s\sigma_{\mu\nu}\gamma_5 d)/2 \\ 
\end{array}
\ee
where $\sigma_{\mu\nu}=[\gamma_\mu, \gamma_\nu]/2$.
In terms of these operators we construct the Fierz eigenbasis,
which we find convenient 
when taking fermion contractions for evaluating the Green
functions in (\ref{eq:wi}),
\be
\begin{array}{ll}
O_0 = \left(VV + AA\right)/2 &(+,+) \\
O_1 = \left(SS+TT+PP\right)/2 &(+,+) \\
O_2 = \left(SS-TT/3+PP\right)/2 &(-,+) \\ 
O_3 = \left(VV-AA\right)/2 + \left(SS-PP\right) &(-,+) \\
O_4 = \left(VV-AA\right)/2 - \left(SS-PP\right) &(+,+) \\ 
\end{array}
\ee
\be
\begin{array}{ll}
O_5 = VA &(+,+) \\
O_6 = SP+T{\tilde T}/2 &(+,-) \\
O_7 = SP-T{\tilde T}/6 &(-,-) \\
\end{array}
\ee
Here the first sign after each equation denotes the Fierz
eigen value and the second the $CPS$\cite{latt88} eigen value. 
We note that the Fierz eigen basis we employ is different from 
the basis chosen by the Rome group\cite{romeBK} based on one-loop
perturbation theory.

The parity odd operators $O_{6,7}$ are $CPS$
odd while $O_5$ is $CPS$ even,  and hence $O_5$ does not mix
with $O_{6,7}$ under renormalization.
Therefore the mixing structure of these operators is given by
\be
\frac{\hat O_{VV+AA}}{2 Z_{VV+AA}}=O_0+z_1O_1+ \cdots
+z_4O_4, 
\label{eq:mix_VV+AA}
\ee
\be
\frac{\hat O_{VA}}{Z_{VA}}= z_5O_5,
\label{eq:mix_VA}
\ee
where $Z_{VV+AA}$ and $Z_{VA}$ are overall renormalization factors.

Let us consider an external state consisting of two $s$ quarks 
and two $d$ quarks, all having an equal momentum $p$.   
Under $\lambda^3$ chiral rotation the Ward identity (\ref{eq:wi}) 
for such an external state takes the following form:
\be
\begin{array}{l}
F_{VV+AA}\equiv \\
-2\rho Z_A^{\rm ext}\langle\sum_xP^a(x)\frac{1}{2}{\hat O}_{VV+AA}(0)
({\tilde s}{\tilde s}
{\tilde {\bar d}}{\tilde {\bar d}})(p)\rangle  \\ 
-\langle\hat O_{VA}(0)
({\tilde s}{\tilde s}
{\tilde {\bar d}}{\tilde {\bar d}})(p)\rangle \\  
-\langle\frac{1}{2}{\hat O}_{VV+AA}(0)
\delta^3({\tilde s}{\tilde s}
{\tilde {\bar d}}{\tilde {\bar d}})(p)\rangle+O(a)=0,  
\end{array}
\ee
\be
\begin{array}{l}
F_{VA}\equiv \\
-2\rho Z_A^{\rm ext}\langle\sum_xP^a(x){\hat O}_{VA}(0)
({\tilde s}{\tilde s}
{\tilde {\bar d}}{\tilde {\bar d}})(p)\rangle  \\ 
-\langle\frac{1}{2}{\hat O}_{VV+AA}(0)
({\tilde s}{\tilde s}
{\tilde {\bar d}}{\tilde {\bar d}})(p)\rangle \\  
-\langle{\hat O}_{VA}(0)
\delta^3({\tilde s}{\tilde s}
{\tilde {\bar d}}{\tilde {\bar d}})(p)\rangle+O(a)=0,  
\end{array}
\ee
where $({\tilde s}{\tilde s}
{\tilde {\bar d}}{\tilde {\bar d}})(p)$ and 
$\delta^3({\tilde s}{\tilde s}
{\tilde {\bar d}}{\tilde {\bar d}})(p)$ represent
${\tilde s}(p){\tilde s}(p)
{\tilde {\bar d}}(p){\tilde {\bar d}}(p)$ and
${\tilde s}(p){\tilde s}(p)
({\tilde {\bar d}}(p)\gamma_5/2){\tilde {\bar d}}(p)$  
$+{\tilde s}(p){\tilde s}(p)
{\tilde {\bar d}}(p)({\tilde {\bar d}}(p)\gamma_5/2)$, respectively.
We obtain the amputated Green functions for 
$F_{VV+AA}$ and $F_{VA}$ by truncating the external quark
propagators according to 
\be
\begin{array}{l}
\Gamma_{VV+AA,VA}\equiv  \\ G_s^{-1}(p)G_s^{-1}(p)F_{VV+AA,VA}
G_{\bar d}^{-1}(p)G_{\bar d}^{-1}(p),
\end{array}
\ee
where $G_q^{-1}$ denotes the inverse quark propagator
with the flavor $q$. 

Let $P_i$ $(i=0,\cdots,7)$ be the projection operator  corresponding
to the four-quark operators in the Fierz eigenbasis 
$O_i$ $(i=0,\cdots,7)$.  For example, we have
\be
P_0^{\alpha\beta\delta\lambda}
=\frac{1}{64}\left[\gamma_\mu^{\alpha\beta}\gamma_\mu^{\delta\lambda}
+(\gamma_\mu\gamma_5)^{\alpha\beta}
(\gamma_\mu\gamma_5)^{\delta\lambda}\right].
\ee
Since QCD conserves parity one can write
\be 
\Gamma_{VV+AA}/Z_{VV+AA}=\Gamma_5P_5,
\ee
\be
\Gamma_{VA}/Z_{VA}=\Gamma_0P_0+\Gamma_1P_1+\cdots+\Gamma_4P_4.
\ee
Expressing ${\hat O_{VV+AA,VA}}$ in  (\ref{eq:wi})
in terms of lattice operators,
we obtain six equations for the five coefficients $z_1,\cdots,z_5$:
\be
\Gamma_i=c^i_0+c^i_1z_1+\cdots +c^i_5z_5=O(a),\, i=0,\cdots, 5\,
\ee 
This gives an overconstrained set of equations, and 
we may choose any five equations to exactly vanish to solve 
for $z_i$: the
remaining equation should automatically be satisfied to
$O(a)$. We choose four equations to be those for 
$i=1,\cdots,4$, since $O_1,\cdots,O_4$ do not
appear in the continuum.
The choice of the fifth equation, $i=0$ or 5,
is more arbitrary. We have checked that either $\Gamma_0=0$
or $\Gamma_5=0$ leads to a
consistent result to $O(a)$ for $z_1,\cdots,z_4$ in the region
$pa\simlt 1$.
In the present analysis we choose $\Gamma_5=0$. 

\begin{table*}[t]
\vspace{-1mm}
\begin{center}
\caption{\label{tab:runpara}Parameters of our simulations. See text for 
details.}
\vspace*{2mm}
\begin{tabular*}{\textwidth}{@{}l@{\extracolsep{\fill}}llll}\hline
    $\beta$       & 5.9   & 6.1   & 6.3   & 6.5 \\ 
\hline
$L^3\times T$     & $24^3\times 64$ & $32^3\times 64$ 
                  & $40^3\times 96$ & $48^3\times 96$ \\
\#conf.           & 300             & 100             
                  & 50              & 24 \\
skip              & 2000 & 2000 & 5000 & 8000 \\
$K$               & 0.15862  & 0.15428  & 0.15131  & 0.14925 \\ 
                  & 0.15785  & 0.15381  & 0.15098  & 0.14901 \\ 
                  & 0.15708  & 0.15333  & 0.15066  & 0.14877 \\ 
                  & 0.15632  & 0.15287  & 0.15034  & 0.14853 \\ 
fitting range ($m_\pi,m_\rho$) 
                  & $13-21$ & $15-25$ & $18-28$ & $21-31$ \\
fitting range ($B_K$)
                  & $19-46$ & $25-40$ & $33-64$ & $36-61$ \\
$K_c$             & $0.15986(3)$ & $0.15502(2)$     
                  & $0.15182(2)$ & $0.14946(3)$ \\ 
$a^{-1}$(GeV)     & 1.95(5)      & 2.65(11)       
                  & 3.41(20)     & 4.30(29) \\ 
$La$(fm)          & 2.4 & 2.4 & 2.3 & 2.2 \\ 
$\alpha_{\overline{\rm MS}}(1/a)$   
                  & 0.1922  & 0.1739  & 0.1596  & 0.1480 \\ 
$m_s a/2=m_d a/2$ & $0.0294(14)$ & $0.0198(16)$     
                  & $0.0144(17)$ & $0.0107(16)$ \\ 
$\delta_{p^2}$    & $1.11$ & $1.11$ & $1.15$ & $1.12$ \\
${p^{*}}^2a^2$    & $0.9595$ & $0.5012$ 
                  & $0.2988$ & $0.2056$ \\ 
\hline
\end{tabular*} 
\end{center}
\vspace{-6mm}
\end{table*}

The overall factor $Z_{VV+AA}$ is determined by the
NPR method\cite{npr}.
We calculate the amputated Green function,
\be
\begin{array}{l}
G_s^{-1}(p)G_s^{-1}(p)
\langle\frac{1}{2}{\hat O}_{VV+AA}(0) 
({\tilde s}{\tilde s}
{\tilde {\bar d}}{\tilde {\bar d}})(p)\rangle  \\ 
\times G_{\bar d}^{-1}(p)G_{\bar d}^{-1}(p) 
={\hat \Gamma}_{VV+AA}(p)P_0+\cdots,
\end{array}
\ee
and impose the following condition,
\be
Z_{VV+AA}(p)Z_q^{-2}(p)
{\hat \Gamma}_{VV+AA}(p)=1,
\ee
where $Z_q(p)$ is the quark wave-function renormalization
factor which is calculated from
\be
Z_q(p)=\frac{{\rm Tr}\left(-i\sum_\mu \gamma_\mu 
{\rm sin}(p_\mu)G_q^{-1}(p)\right)}
{4\sum_\mu {\rm sin}^2(p_\mu)}.
\ee

We convert the matrix elements on the lattice  into  those of the
$\overline{\rm MS}$ scheme in the continuum with naive dimensional
regularization (NDR) and renormalized at the scale
$\mu=2$GeV\cite{z_bk}:
\be
B_K({\rm NDR},\mu)=B_K(p,1/a)
\label{eq:Z-factor}
\ee
\vspace{-4mm}
\[
\times\left[1+\frac{\alpha_s(\mu)}{4\pi}\left(-4\, {\rm
log}\left(\frac{\mu}{p}\right)
-\frac{14}{3}+8\log 2 \right)\right],
\]
where 
\be
B_K(p,1/a)=\frac{\langle
{\bar K}^0 \vert \hat O_{VV+AA} \vert K^0 \rangle}
{\frac{8}{3}\vert \langle 0 \vert {\hat A} 
\vert K^0 \rangle \vert^2}
\ee
with $p$ the momentum at which the mixing coefficients are evaluated.
The axial vector current in the denominator is given by
${\hat A}=Z_A\bar s\gamma_4\gamma_5 d$ 
with $Z_A$ the renormalization factor determined by the NPR
method.

For comparative purpose we also calculate $B_K$ with 
perturbative  mixing coefficients, for which we use the one-loop
expression in Ref.~\cite{pt} after 
applying a finite correction in conversion to 
the NDR scheme together with  
the tadpole improvement with 
$\alpha_{\overline{\rm MS}}(1/a)$.

Let us remark here that the equations obtained in the NPR
method \cite{romeBK} corresponds to $\Gamma_i=0$ for $i=1, \cdots, 4$ 
in which the contributions of the first and the third term in the Ward 
identity (\ref{eq:wi}) are dropped.  In particular the NPR method 
neglects the quark mass contributions coming from the first
term of (\ref{eq:wi}).  As may be expected from this, the NPR method 
is equivalent to the Ward identities in the limit of large external
virtualities\cite{npr,romeBK}.

\section{Parameters of numerical simulation}
  
Our calculations are made with the Wilson quark action and the plaquette 
action at $\beta=5.9-6.5$ in quenched QCD. 
Table~\ref{tab:runpara} summarizes our run parameters. 
Gauge configurations are generated with the 5-hit pseudo heat-bath
algorithm.  At each value of $\beta$ 
four values of the hopping parameter $K$ are adopted such that the physical
point for the $K$ meson can be interpolated.  
The critical hopping parameter $K_c$ is determined by
extrapolating results for $m_\pi^2$ for the four hopping parameters 
linearly in $1/2K$ to $m_\pi^2=0$.
We take the down and strange quarks to be degenerate.  
The value of half the strange quark mass $m_s a/2$ is then estimated 
from $m_K/m_\rho=0.648$.

The physical size of lattice is chosen to be approximately 
constant at $La\approx 2.4$fm where the lattice spacing 
is determined from $m_\rho=770$MeV.
To calculate the perturbative renormalization factors, 
we employ the strong coupling constant at the scale $1/a$ 
in the $\overline{\rm MS}$ scheme, evaluated by a two-loop 
renormalization group running starting from   
$g^2_{\overline{\rm MS}}(\pi/a)=
P_{\rm av}/g^2_{latt}+0.0246$ with $P_{\rm av}$ the averaged value of the
plaquette.

The renormalization factors $Z_{VV+AA}$, $Z_A$ and the mixing
coefficients $z_i$ $(i=1,\cdots,5)$ are calculated for a set of 
external quark momenta $p^{(i)}$ $(i=1,\cdots,\sim 40)$, which is
chosen  recursively according to the condition
that the $(i+1)$-th momentum $p^{(i+1)}a$ is the minimum number
satisfying 
${p^{(i+1)}}^2a^2\ge \delta_{p^2}\cdot {p^{(i)}}^2a^2$
with $p^{(1)}a=2\pi/T$ where $T$ denotes the temporal lattice size.
We employ $p^*\approx 2$GeV among $p^{(i)}$ for the analysis of 
$B_K$.
Errors are estimated by the single elimination 
jackknife method for all measured quantities
throughout this work.

\section{Calculational procedure}

Our calculations are carried out in two steps. We first calculate
$Z_{VV+AA}$, $Z_A$ and $z_i$ using the quark Green functions 
having finite space-time momenta. For this purpose 
quark propagators are solved in the Landau gauge
for the point source located at the origin
with the periodic boundary condition imposed in all four directions.
For calculating the first term of the Ward identity (\ref{eq:wi}), 
we use the source method\cite{source} 
to insert the pseudoscalar density.
The point source quark propagators are also used for calculating 
$\pi$ and $\rho$ propagators and to extract their masses from them.

The $B_K$ parameter is extracted from the ratio 
\ben
R(t^\prime)&=&\frac{\langle O_{{\bar K}^0}(T){\hat
O}_{VV+AA}(t^{\prime})O_{K^0}^\dagger(1)\rangle }{\frac{8}{3} 
\langle O_{{\bar K}^0}(T){\hat A}(t^{\prime})\rangle
\langle {\hat A}(t^{\prime})O_{K^0}^\dagger(1)\rangle} 
\label{eq:ratio} \\
&\longrightarrow& B_K(p^*,1/a)/L^3\;\;{\rm for}\;\; 1\ll t^\prime \ll T,\nonumber
\een
where various operators are defined by
$O_{K^0}(t)=\sum_{\vec x}{\bar s}(\vec{x},t)\gamma_5 d(\vec{ x},t)$,
$O_{\bar K^0}(t)$$=$$\sum_{\vec x}{\bar d}(\vec{x},t)\gamma_5 s(\vec{ x},t)$,
${\hat O}_{VV+AA}(t)$$
=$$\sum_{i=0}^{4}\sum_{\vec x}Z_{VV+AA} z_i O_i(\vec{ x},t)$ and
${\hat A}(t)$$=$$\sum_{\vec x} Z_A {\bar s}(\vec{x},t)
\gamma_4 \gamma_5 d(\vec{ x},t)$.
The contribution of the operators $O_i$
$(i=0,\cdots,4)$ to $B_K(p^*,1/a)$ can be obtained similarly 
from 
\be
R^i(t^\prime)=\frac{\langle O_{{\bar K}^0}(T)
{\hat O}_i(t^\prime)
O_{K^0}^\dagger(1)\rangle }{\frac{8}{3} 
\langle O_{{\bar K}^0}(T){\hat A}(t^{\prime})\rangle
\langle {\hat A}(t^{\prime})O_{K^0}^\dagger(1)\rangle},  
\label{eq:ratio_i}
\ee 
where ${\hat O_i}(t^\prime)=
\sum_{\vec x}Z_{VV+AA} z_i O_i(\vec{ x},t^\prime)$.
For calculation of these ratios we solve 
quark propagators without gauge fixing employing wall
sources placed at the edges of lattice where the Dirichlet boundary
condition is imposed in the time
direction. We obtain $B_K$ at $m_s/2$ by quadratically 
interpolating the data at the four hopping parameters.

\section{Results for mixing coefficients}

\begin{figure}[t]
\centering{
\hskip -0.0cm
\psfig{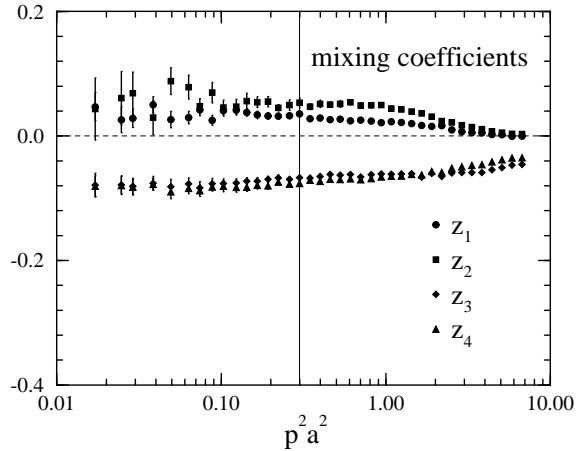}
\vskip -10mm  }
\caption{Mixing coefficients $z_1,\cdots,z_4$ 
plotted as a function of external momentum squared $(pa)^2$   
for $K=0.15034$ at $\beta=6.3$. Vertical line corresponds to 
$p^*\approx 2$ GeV.} 
\label{fig:zmix_63_ours}
\vspace{-6mm}
\end{figure}

\begin{figure}[t]
\centering{
\hskip -0.0cm
\psfig{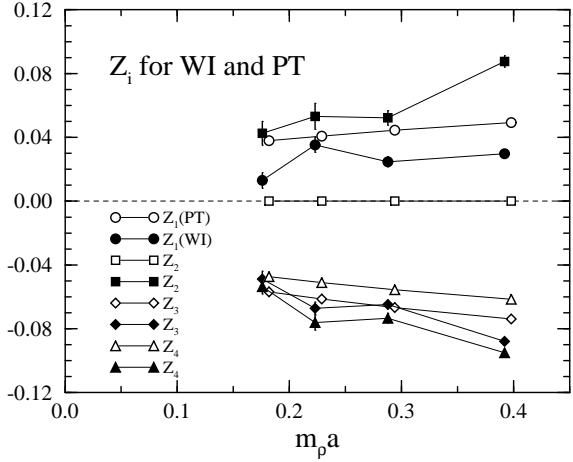}
\vskip -10mm  }
\caption{Comparison of the mixing coefficients
$z_1,\cdots,z_4$ evaluated at ${p^{*}}\approx 2$ GeV 
using the  Ward identity (WI; solid symbols) 
and perturbative (PT; open symbols) methods. 
The coefficients are plotted as a function of $m_\rho a$.}  
\label{fig:zmix_beta}
\vspace{-6mm}
\end{figure}

\begin{figure}[t]
\centering{
\hskip -0.0cm
\psfig{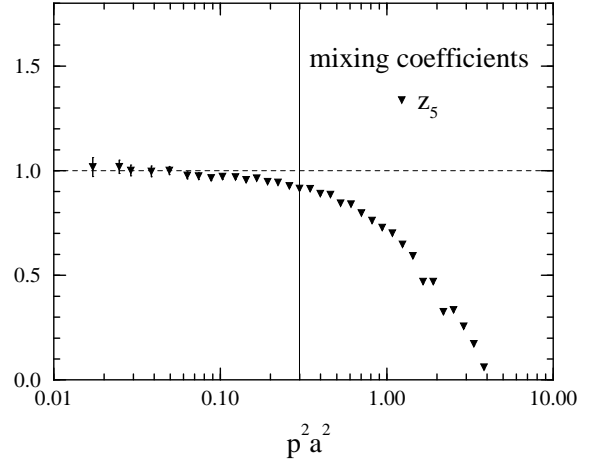}
\vskip -9mm  }
\caption{Same as Fig.~\protect{\ref{fig:zmix_63_ours}}
for mixing coefficient $z_5$.} 
\label{fig:zmix_5_63}
\vspace{-6mm}
\end{figure}

\begin{figure}[t]
\centering{
\hskip -0.0cm
\psfig{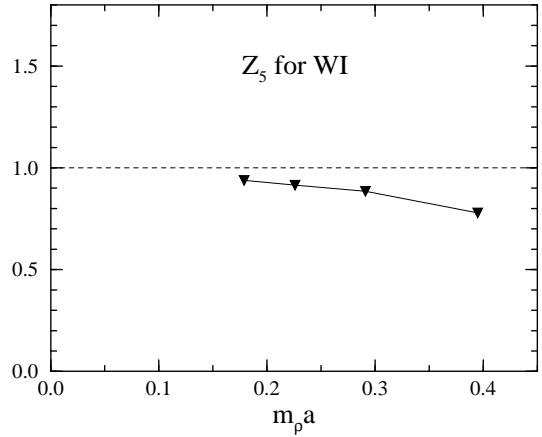}
\vskip -6mm  }
\caption{Mixing coefficient $z_5$ evaluated 
at ${p^{*}}\approx 2$ GeV using the  Ward identity method 
as a function of $m_\rho a$.}  
\label{fig:zmix_5_beta}
\vspace{-6mm}
\end{figure}

In Fig.~\ref{fig:zmix_63_ours} we plot 
a typical result for the mixing coefficients $z_i$ $(i=1,\cdots,4)$ 
as a function of the external quark momenta for the case of 
$K=0.15034$ at $\beta=6.3$. The mixing coefficients 
shows only weak dependence over a wide momentum range  
$0.02 \simlt p^2a^2 \simlt 1.0$, albeit $z_1$ and
$z_2$ have large errors in the small momentum 
region $ p^2a^2 \simlt 0.1$. 
This enables us to evaluate the mixing coefficients with small errors 
at the scale $p^{*}\approx 2$GeV, which
always falls within the range of a plateau 
for our runs at $\beta=5.9-6.5$.

At the workshop Talevi\cite{talevi} presented a reanalysis of the 
results of the Rome group for the mixing coefficients in the 
Fierz eigen basis, reporting that the momentum dependence in this 
basis is similar to that of our results in Fig.~\ref{fig:zmix_63_ours}.  
Their simulations are made at $\beta=6.2$ with the Clover action.
For a more detailed comparison, a parallel analysis with the NPR and 
Ward identity methods employing the same quark action 
on the same set of configurations would be desirable.
 
In Fig.~\ref{fig:zmix_beta} we compare the mixing
coefficients $z_i$ $(i=1,\cdots,4)$ evaluated 
at the scale $p^{*}$ (filled symbols)
with the perturbative values obtained with 
$\alpha_{\overline{\rm MS}}(1/a)$ (open symbols) as a function of lattice 
spacing.
We observe that the $a$ dependence for the mixing
coefficients determined from the Ward identities is steeper compared
to that for the perturbative estimates.
The magnitude of each mixing coefficient for the Ward
identity method varies nearly in
proportion to $a$, which reduces by $50\%$ between
$m_\rho a\approx 0.4$ and $m_\rho a\approx 0.2$.

We remark that a large value of $z_2$ determined by the Ward identities 
sharply contrasts with the one-loop perturbative result $z_2=0$.  
For the other coefficients, the perturbative
results agree with the non-perturbative ones in sign and
rough orders of magnitude. However, they differ in quantitative detail.
We find that the magnitudes of $z_4$ are larger than those of 
$z_3$ for all values of $\beta$, which is contrary to the
perturbative result.

For our study of the $B$ parameter the mixing coefficient $z_5$
for the parity odd operator ${\hat O_{VA}}$ is not directly relevant. 
For completeness we plot a typical result in Fig.~\ref{fig:zmix_5_63}.
The data shows a scale dependence which is stronger than those of 
$z_i$ $(i=1,\cdots,4)$ for parity even operators toward large momenta.
We do not find this to be particularly alarming since $z_5$ 
evaluated for a fixed physical scale $p^*$ approaches unity
toward the continuum limit as shown in Fig.~\ref{fig:zmix_5_beta}.

\section{Chiral behavior}

\begin{figure}[t]
\centering{
\hskip -0.0cm
\psfig{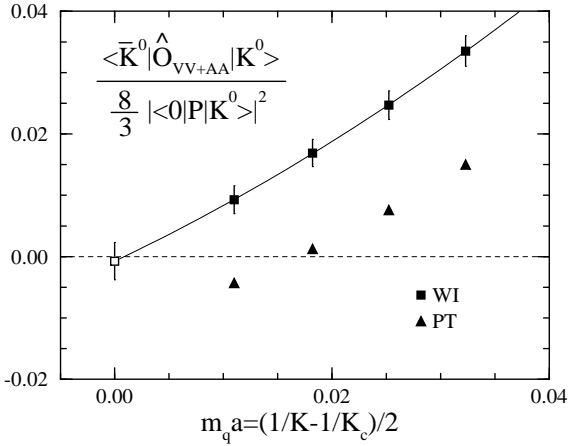}
\vskip -10mm  }
\caption{Test of the chiral behavior of  
$\langle\bar K^0|\hat O_{VV+AA}|K^0\rangle/
(8/3)/|\langle 0|P|K^0\rangle|^2$ 
for the WI and PT methods at $\beta=6.3$.
The solid curve is a quadratic extrapolation to the chiral limit.} 
\label{fig:bk_p_63}
\vspace{-6mm}
\end{figure}

\begin{figure}[t]
\centering{
\hskip -0.0cm
\psfig{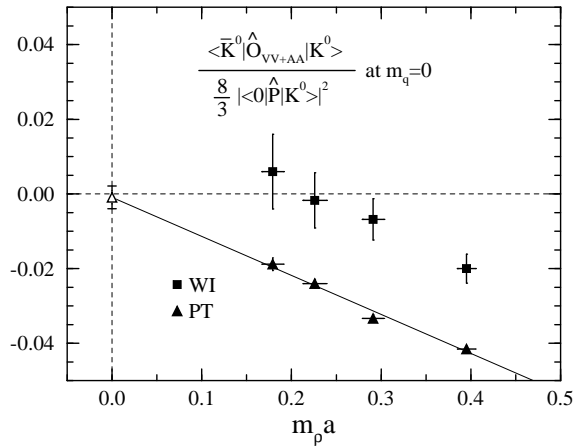}
\vskip -10mm  }
\caption{$\langle\bar K^0|\hat O_{VV+AA}|K^0\rangle/(8/3)/
|\langle 0|\hat P|K^0\rangle|^2$ at $m_q=0$
for the WI and PT methods as a function of $a$.  
The operators are renormalized at 2 GeV in the NDR scheme.
For both methods we use the same $\hat P$ perturbatively corrected
with the tadpole improvement.
The solid line is a linear extrapolation to the continuum 
limit.}
\label{fig:bk_p_cl}
\vspace{-6mm}
\end{figure}

Let us examine the chiral property of the operator 
$\hat O_{VV+AA}$.
In Fig.~\ref{fig:bk_p_63} we show the chiral behavior of the ratio
$\langle{\bar K}^0 \vert \hat O_{VV+AA} \vert K^0
\rangle/(8/3)$  $/
\vert \langle 0 \vert P \vert K^0 \rangle \vert^2$ at $\beta=6.3$,
where WI stands for our method using chiral Ward identities and PT for  
tadpole-improved one-loop perturbation theory.
The solid line represents a quadratic extrapolation of the Ward identity 
result in the bare quark mass $m_q a=(1/K-1/K_c)/2$.
The extrapolated value at $m_q=0$ is consistent with zero, 
demonstrating a significant improvement of the chiral behavior 
compared to the perturbative result plotted with triangles.  

We plot in Fig.~\ref{fig:bk_p_cl} the values of the ratio extrapolated to 
$m_q=0$ as a function of lattice spacing, 
where the pseudoscalar density $\hat P$ in the denominator 
is renormalized perturbatively for both WI and PT cases
(numerical results are given in Table~\ref{tab:bk} below).
The ratio for the Ward identity method  becomes consistent with zero at
the lattice spacing $m_\rho a\simlt 0.3 (a\simlt 0.08$fm).

In the perturbative approach
with the one-loop mixing coefficients, chiral breaking
effects are expected to appear as terms of $O(g^4)$ and
$O(a)$ for the Wilson quark action.  
A roughly linear behavior of our results for the perturbative method 
is consistent with the presence of an $O(a)$ term. 
Making a linear extrapolation to the continuum limit $a\to 0$, we observe that 
the chiral behavior is recovered.  
This may suggest that the $O(g^4)$ terms in the mixing 
coefficients left out in the one-loop treatment are actually small.

\section{Results for $B_K$}

\begin{figure}[t]
\centering{
\hskip -0.0cm
\psfig{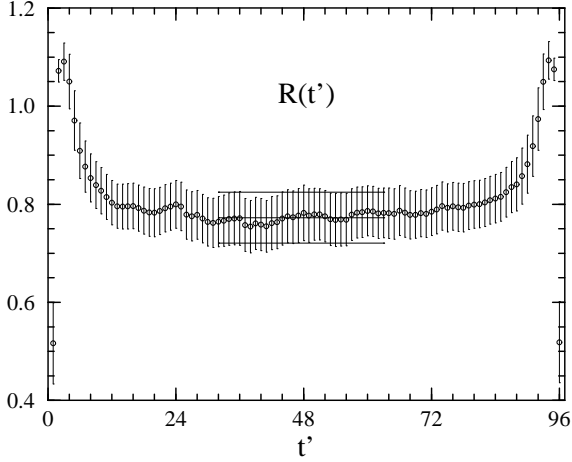}
\vskip -10mm  }
\caption{Ratio $R(t^\prime)$
for $K=0.15034$ at $\beta=6.3$. Solid lines denote 
the fitted result and a one standard deviation error band.}
\label{fig:bk_t_63}
\vspace{-3mm}
\end{figure}

\begin{figure}[t]
\centering{
\hskip -0.0cm
\psfig{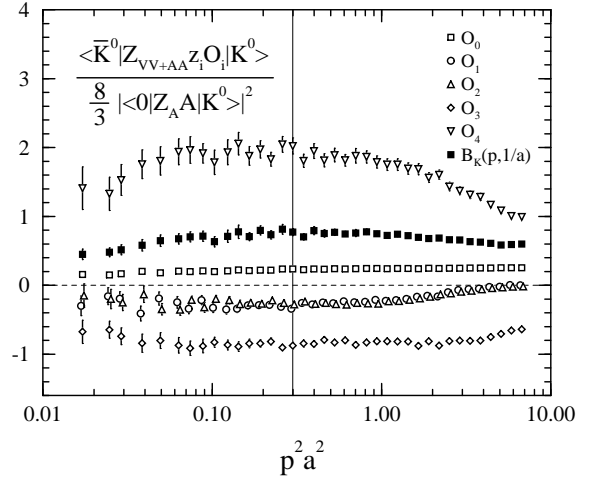}
\vskip -10mm  }
\caption{Contribution of the operators 
$O_i$ $(i=0,\cdots,4)$ to $B_K(p,1/a)$ with $z_i$
determined by the WI method
for $K=0.15034$ at $\beta=6.3$. Vertical line corresponds to 
$p^*\approx 2$ GeV.}
\label{fig:bk_all_63}
\vspace{-3mm}
\end{figure}

\begin{figure}[t]
\centering{
\hskip -0.0cm
\psfig{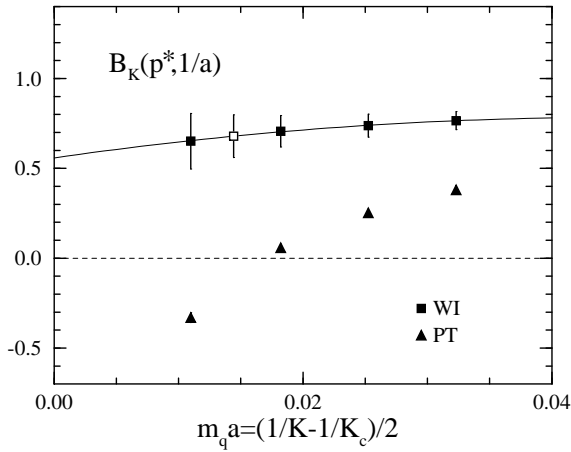}
\vskip -10mm  }
\caption{Quark mass dependences of $B_K(p^*,1/a)$
for the WI and PT method at $\beta=6.3$.
The open symbol is a quadratical interpolation of the data 
to $m_s/2$.}
\label{fig:bk_a_63}
\vspace{-6mm}
\end{figure}

\begin{figure}[t]
\centering{
\hskip -0.0cm
\psfig{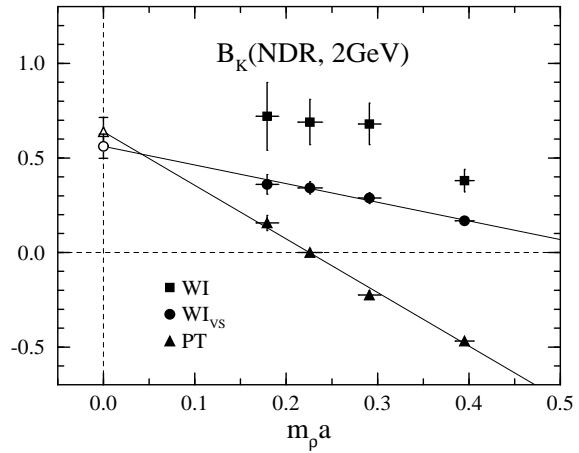}
\vskip -10mm  }
\caption{$B_K($NDR, 2GeV) plotted as a function 
of $a$ for the WI, WI$_{\rm VS}$ and PT methods. 
The solid lines show linear 
extrapolations to the continuum limit.}
\label{fig:bk_a_cl}
\vspace{-6mm}
\end{figure}

We now turn to the calculation of $B_K({\rm NDR},$ $2{\rm GeV})$.
In Fig.~\ref{fig:bk_t_63} we present the ratio $R(t^\prime)$ 
defined in  (\ref{eq:ratio}) using the mixing coefficients 
determined from the Ward identities
for $K=0.15034$ at $\beta=6.3$. A good
plateau is observed in the range $20\simlt t^\prime \simlt 70$.
We make a global fit of the ratio $R(t^\prime)$ 
to a constant over $33\le t^\prime \le 64$ for this data set.
The three horizontal lines denote the central value of the fit and 
a one standard deviation error band.
We note that the error of the fitted
result is roughly  equal in magnitude to those of the ratio over the 
fitted range, while we would usually expect a smaller error for the 
fitted result.
This is because the error of 
the ratio $R(t^\prime)$ is
governed by those of the mixing coefficients 
$z_i$ $(i=1,\cdots,4)$.

In Fig.~\ref{fig:bk_all_63}
a representative result for 
the contribution of each operator 
$O_i$ $(i=0,\cdots,4)$ to $B_K(p,1/a)$ is plotted 
as a function of the external quark momenta, 
which is obtained by fitting the
ratio $R^i(t^\prime)$ $(i=0,\cdots,4)$ of (\ref{eq:ratio_i}) 
with a constant over the same fitting range 
as for $R(t^\prime)$. 
The contributions are nearly independent of the
external quark momentum in the 
range $0.05\simlt p^2 a^2 \simlt 1.0$
as is expected from the weak scale dependence of the
mixing coefficients $z_i$ 
shown in Fig.~\ref{fig:zmix_63_ours}.	 
A decrease in magnitude of the contributions 
in the small momentum region $p^2 a^2 \simlt 0.05$
originates from the scale dependence of the overall
renormalization factor $Z_{VV+AA}/Z_A^2$ estimated non-perturbatively.  

An important observation is that 
the value of $B_K(p,1/a)$ is a result of a 
large cancellation between the amplitudes of $z_iO_i$ where 
each amplitude of the mixing operator $z_i O_i$ $(i=1,\cdots,4)$ is
comparable to or larger than that of $O_0$.  
This is the reason why calculations of $B_K$ with the Wilson quark action 
is difficult.

We plot the quark mass dependence of $B_K(p^*,1/a)$ at $\beta=6.3$ in 
Fig.~\ref{fig:bk_a_63}.
We observe that the results for the PT method  
seems to diverge toward the chiral limit, while those for
the WI method stays finite.
Interpolating the data for the four hopping parameters 
quadratically to $m_s/2$ we obtain the value of $B_K(p^*,1/a)$
at the physical point.

Let us note that the perturbative results have quite small errors 
compared to those of the Ward identity method.  This is because 
the mixing coefficients are definitely given for the perturbative method.
For a precise determination of $B_K$ with the Ward identity method 
it is of great importance to reduce the errors of the mixing coefficients.
For this purpose methods have to be devised to effectively compute 
quark propagators for a large set of momenta with good precision.

\begin{table*}[t]
\vspace{-1mm}
\begin{center}
\caption{\label{tab:bk}
Results for $B_K($NDR, 2GeV) for WI, WI$_{\rm VS}$ and PT methods 
as a function of $\beta$. Ratio 
$\langle{\bar K}^0 \vert \hat O_{VV+AA} \vert K^0 \rangle /(8/3)/
\vert \langle 0 \vert {\hat P} \vert K^0 \rangle \vert^2$ 
in the chiral limit is also given.}
\vspace*{2mm}
\begin{tabular*}{\textwidth}{@{}l@{\extracolsep{\fill}}llllll}\hline
    $\beta$       & & 5.9   & 6.1   & 6.3   & 6.5   & $a=0$ \\ 
\hline
$B_K($NDR, 2GeV) 
             & WI     
                  & $+0.38(6)$    & $+0.68(11)$ 
                  & $+0.69(12)$   & $+0.72(18)$   & $$ \\
             & WI$_{\rm VS}$     
                  & $+0.168(20)$  & $+0.288(29)$  
                  & $+0.342(33)$  & $+0.360(52)$  & $+0.562(64)$ \\
             & PT     
                  & $-0.468(14)$  & $-0.225(22)$   
                  & $-0.000(21)$  & $+0.156(40)$  & $+0.639(76)$ \\
\hline
$\left.\frac{\langle {\bar K}^0 \vert \hat O_{VV+AA} \vert K^0 \rangle }
{\frac{8}{3}\vert \langle 0 \vert {\hat P} 
\vert K^0 \rangle \vert^2}\right|_{m_q=0}$ 
             & WI
                  & $-0.0200(39)$ & $-0.0068(55)$   
                  & $-0.0017(74)$ & $+0.006(10)$  & $$ \\
             & PT
                  & $-0.0415(8)$  & $-0.0333(10)$   
                  & $-0.0240(12)$ & $-0.0188(17)$ & $-0.0009(31)$ \\
\hline
\end{tabular*} 
\end{center}
\vspace{0mm}
\end{table*}

Our final results for $B_K({\rm NDR},2{\rm GeV})$ obtained with 
 (\ref{eq:Z-factor}) are presented in Fig.~\ref{fig:bk_a_cl} as 
a function of lattice spacing.  The numerical values are listed 
in Table~\ref{tab:bk}.
The method based on the Ward
identity (WI) gives a value convergent from a 
lattice spacing of $m_\rho a\approx 0.3$ ($B_K\sim0.6-0.8$). 
The large error, however, hinders us from making an extrapolation to 
the continuum limit.  Since the origin of the large error is traced to 
that of the mixing coefficients,
we develop an alternative method, which we refer to as 
WI$_{\rm VS}$, in which the denominator of the
ratio for extracting $B_K$ is estimated with the vacuum
saturation of $\hat O_{VV+AA}$ constructed by the WI method:
\be
B_K^{\rm VS}(p^*,1/a)=\frac{\langle
{\bar K}^0 \vert \hat O_{VV+AA} \vert K^0 \rangle}
{Z_A^2 \sum_{i=0}^4 z_i \langle{\bar K}^0 \vert O_i 
\vert K^0 \rangle_{\rm VS}}.
\ee
In this case the fluctuations in the numerator, mainly due
to those of the mixing coefficients $z_i$, are largely
canceled by those in the denominator, and the resulting
error in $B_K$ is substantially reduced as apparent in 
Fig.~\ref{fig:bk_a_cl}. 
The cost is that the denominator of $B_K^{\rm VS}(p^*,1/a)$
contains the contributions of the pseudoscalar density $\langle
{\bar K}^0 \vert P\vert 0\rangle \langle 0\vert P\vert K^0\rangle$
besides those of the axial vector current 
$\langle
{\bar K}^0 \vert {\hat A}\vert 0\rangle \langle 0\vert {\hat 
A}\vert K^0\rangle$,
due to which the correct chiral behavior 
of the denominator is not respected  
at a finite lattice spacing.
While WI and WI$_{\rm VS}$ methods give different results at
a finite lattice spacing, the discrepancy is expected 
to vanish in the continuum limit.
A linear extrapolation in $a$ of the WI$_{\rm VS}$ results yields 
$B_K({\rm NDR},2{\rm GeV})=0.562(64)$, which we take as the best value
in the present work. This value is consistent
with a recent JLQCD result with the Kogut-Susskind
action,
$B_K({\rm NDR},2{\rm GeV})=0.587(7)(17)$\cite{saoki}.

Intriguing in Fig.~\ref{fig:bk_a_cl} is that the perturbative calculation (PT),
which gives completely ``wrong'' values at $a\ne0$,
also yields the correct result for $B_K$, when extrapolated
to the continuum limit $a=0$. 
This is a long extrapolation from negative to positive, 
but the linearly extrapolated value $B_K$(NDR, 2GeV)=0.639(76) is consistent 
with those obtained with the WI or WI$_{\rm VS}$ method.
We note that a long extrapolation may bring a large error in the 
extrapolated value.

Finally we mention possible sources of systematic errors in our results 
from quenching effects and uncertainties for Gribov 
copies in the Landau gauge. 
With the Kogut-Susskind quark action it has been observed that the error
due to quenched approximation is small \cite{OSU}. Whether this
is supported by calculations with Wilson action we must defer to future 
studies. 
For the Gribov problem we only quote an earlier study\cite{gribov} 
which suggests that ambiguities in the choice of the Gribov copies 
induce only small uncertainties comparable to typical statistical errors 
in current numerical simulations.  

\section{Conclusions}
 
Our analysis of $B_K$ demonstrates the effectiveness 
of the method of chiral Ward identities for constructing the $\Delta
s=2$ operator with the correct chiral property.  We have shown that both
Wilson and Kogut-Susskind actions give virtually the
identical answer for $B_K$ in their
continuum limit. 
We may hope that further improvement of our simulations, 
especially the reduction of the errors for the mixing
coefficients, leads to a precise
determination of $B_K$ with the Wilson quark action.  
The application of this method
for calculations of $B_B$ is straightforward. 


\vspace{-2mm} 
\section*{Acknowledgements}

This work is supported by the Supercomputer Project (No.1) of 
High Energy Accelerator Research Organization(KEK), and also 
in part by the Grants-in-Aid of the Ministry of Education 
(Nos. 08640349, 08640350, 08640404, 08740189, 08740221).

\end{document}